\documentclass[showpacs,preprintnumbers,amsmath,prl,graphicx,amssymb,superscriptaddress,twocolumn]{revtex4}
\usepackage{graphicx}
\usepackage{dcolumn}
\usepackage{bm}

\renewcommand{\today}{\number\day\space\ifcase\month\or January\or 
 February\or March\or April\or May\or June\or July\or August\or 
 September\or October\or November\or December\fi\space\number\year}

\begin{document}
\title{Experimental constraints on a dark matter origin\\ for the 
DAMA annual modulation effect}

\newcommand{\pnnl}{Pacific Northwest National Laboratory, Richland,
WA 99352, USA}
\newcommand{\uc}{Kavli Institute for Cosmological Physics and Enrico
Fermi Institute, University of Chicago, Chicago, IL 60637, USA}
\newcommand{\madrid}{Departamento de F\'{i}sica Te\'{o}rica C-XI \&
Instituto de F\'{i}sica Te\'{o}rica UAM-CSIC, Universidad
Aut\'{o}noma de Madrid, Cantoblanco, E-28049 Madrid, Spain}
\newcommand{\canberra}{CANBERRA Industries, Meriden, CT 06450, USA}
\newcommand{\anl}{Argonne National Laboratory, Argonne, IL 60439, USA}
\newcommand{\snl}{Sandia National Laboratories, Livermore, CA 94550,
USA}
\newcommand{\uw}{Center for Experimental Nuclear Physics and
Astrophysics and Department of Physics, University of Washington,
Seattle, WA 98195}


\affiliation{\pnnl}
\affiliation{\uc}
\affiliation{\madrid}
\affiliation{\canberra}
\affiliation{\anl}
\affiliation{\snl}
\affiliation{\uw}
														
\author{C.E.~Aalseth}\affiliation{\pnnl}
\author{P.S.~Barbeau}\affiliation{\uc}
\author{D.G.~Cerde\~no}\affiliation{\madrid}
\author{J.~Colaresi}\affiliation{\canberra}
\author{J.I.~Collar$^{*}$}\affiliation{\uc}
\author{P.~de Lurgio}\affiliation{\anl}
\author{G.~Drake}\affiliation{\anl}
\author{J.E.~Fast}\affiliation{\pnnl}
\author{C.H.~Greenberg}\affiliation{\uc}
\author{T.W.~Hossbach}\affiliation{\pnnl}
\author{J.D.~Kephart}\affiliation{\pnnl}
\author{M.G.~Marino}\affiliation{\uw}
\author{H.S.~Miley}\affiliation{\pnnl}
\author{J.L.~Orrell}\affiliation{\pnnl}
\author{D.~Reyna}\affiliation{\snl}
\author{R.G.H.~Robertson}\affiliation{\uw}
\author{R.L.~Talaga}\affiliation{\anl}
\author{O.~Tench}\affiliation{\canberra}
\author{T.D.~Van Wechel}\affiliation{\uw}
\author{J.F.~Wilkerson}\affiliation{\uw}
\author{K.M.~Yocum}\affiliation{\canberra}

\collaboration{CoGeNT Collaboration}
\noaffiliation


\begin{abstract}
A claim for 
evidence of  
dark matter interactions in the DAMA experiment has been recently reinforced. 
We employ a new type of germanium detector to
conclusively 
rule out a standard isothermal galactic halo of Weakly 
Interacting Massive Particles (WIMPs) 
as the explanation for the annual modulation effect leading to the 
claim. 
Bounds are similarly imposed on a suggestion that dark  
pseudoscalars might
lead to the
effect. We describe the sensitivity to light dark 
matter particles achievable with our device, in 
particular to Next-to-Minimal Supersymmetric 
Model candidates.
\end{abstract}

\pacs{85.30.-z, 95.35.+d, 95.55.Vj, 14.80.Mz\\
$^{*}$~Corresponding author. E-mail: collar@uchicago.edu}

\maketitle
The DAMA and DAMA/LIBRA \cite{damalibra} experiments 
have accumulated a combined 0.82 ton-years 
of NaI[Tl] exposure to putative dark matter particles, 
substantially exceeding that from  
any other dedicated search. The newer 
DAMA/LIBRA array 
features a larger target mass and
an improved internal 
radiopurity. The first DAMA/LIBRA dataset has
confirmed the evidence for an annual 
modulation in the few keV portion of the spectrum
\cite{damaclaim}, 
an effect previously observed in DAMA. The observed modulation has
all the characteristics 
(amplitude, phase, period)
expected \cite{annual} from the motion of an Earth-bound laboratory
through a 
standard isothermal halo composed of WIMPs. The statistical significance of the modulation has reached 
8.2 sigma. No other explanation has been found 
yet, prompting the DAMA collaboration to claim the effect is due to 
dark matter interactions.

Competing dark matter searches have been able to exclude most of the 
phase space (nuclear scattering cross section {\frenchspacing vs.
WIMP} mass) 
available as an explanation for this time-modulated signal. 
However, as a result of insufficiently-low energy 
thresholds in those detectors, it has been proposed
\cite{sdgraciela,sigraciela}
that light WIMPs of less than $\sim\!10~$GeV/c$^{2}$ could cause 
the observed modulation while avoiding existing experimental 
constraints.
This hypothesis has been recently ruled out
by COUPP \cite{science},
for those cases where WIMP-nucleus scattering is mediated by a
spin-dependent 
coupling \cite{sdgraciela}. The  
results presented here exclude 
the remaining spin-independent phase space \cite{sigraciela}. These new limits
effectively preclude a standard WIMP halo as a viable explanation 
for the DAMA observations. 

A new type of germanium radiation detector with an unprecedented
combination of crystal mass 
and sensitivity to sub-keV signals has been described in 
\cite{jcap}.
These detectors provide
significant 
improvements over conventional coaxial  
designs ({\frenchspacing Fig. 1}). Details on the modifications leading to this
performance, as well 
as a description of the applications for this device,
can be found in \cite{jcap}. We refer to this design as 
a p-type point contact (PPC) germanium detector (HPGe).

\begin{figure}
\includegraphics[width=7.5cm]{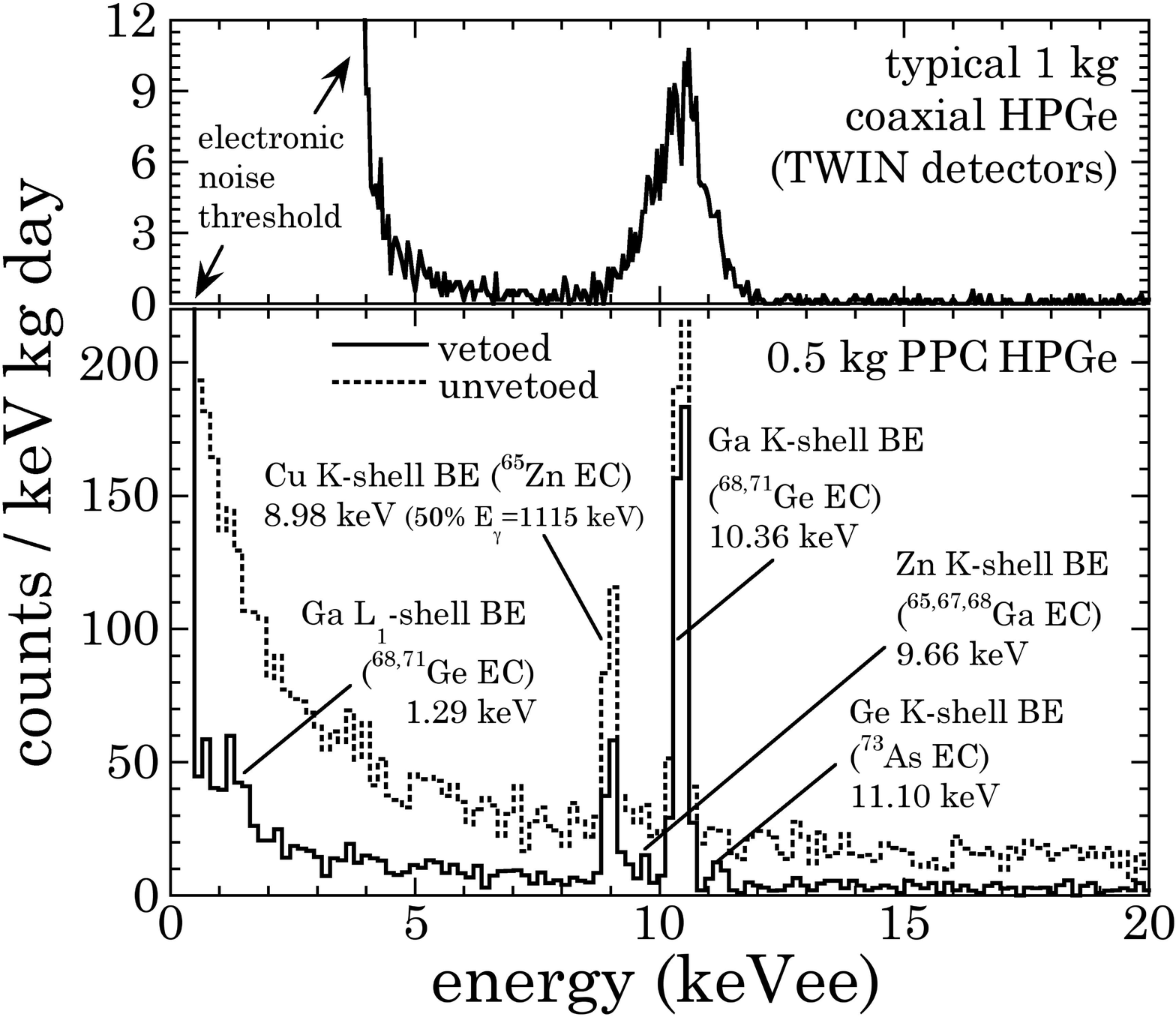}
\caption{\label{fig:epsart} Improvements in threshold and 
resolution in a PPC design (bottom), compared to a 
typical coaxial HPGe \protect\cite{twin} (top). 
Cosmogenic peaks are clearly resolved in the PPC spectrum. BE stands 
for binding energy, EC for electron capture.}
\end{figure}

Several PPCs have been successfully built since the description of 
the first prototype, most within the {\sc{Majo\-ra\-na}} 
collaboration \cite{majorana}. The dataset utilized here comes from 
tests of the first prototype in a shallow underground location (330 
m.w.e., a pumping station part of the Tunnel And Reservoir Plan of 
the city of Chicago). 
While the 
results obtained already impose constraints on the possible dark 
matter
origins of the DAMA anomaly, it is expected 
that ongoing cryostat improvements, a 
longer exposure (8.4 kg-days here) and operation in a deeper 
laboratory should dramatically
improve the dark matter sensitivity of the device. The potential
reach of this method is 
discussed in more detail below.

Listing from the innermost to the outermost components, 
the shield installed around the detector was: 
{\it i}) a 10 cm-thick, low-background NaI[Tl] anti-Compton veto,
{\it ii}) 5 
cm of low-background lead, {\it iii})
15 
cm of standard lead, {\it iv}) 0.5 
cm of borated neutron absorber, {\it v}) a $>$99.9\% efficient muon veto,  {\it vi}) 30 
cm of neutron moderator, and {\it vii}) a 
low-efficiency large-area external muon veto. 
{\frenchspacing Fig. 1} shows the magnitude of the active background 
rejection. The rate of random coincidences 
between 
PPC 
and active element events, measured with a pulser, was 
$\sim$18\%. The low-energy dataset 
used for dark matter limit extraction (inset, {\frenchspacing Fig.
2}) has been 
corrected to account for these. 

\begin{figure}
\includegraphics[width=8cm]{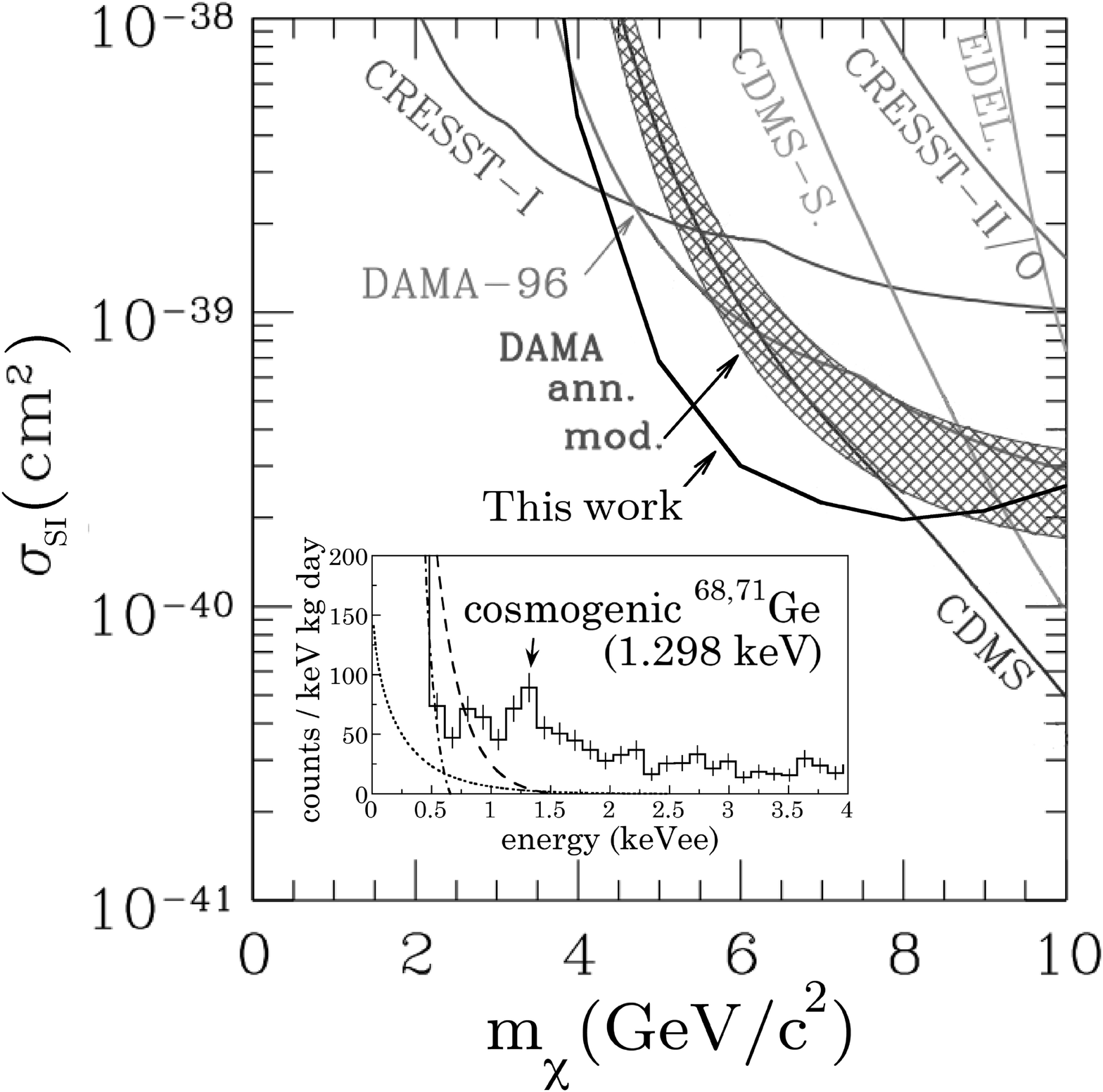}
\caption{\label{fig:epsart}  Parameter space region (cross-hatched)
able to 
explain the DAMA modulation via spin-independent 
couplings from an isothermal light-WIMP halo 
\protect\cite{sigraciela}. 
Lines delimit the coupling 
($\sigma_{SI}$) vs. WIMP mass ($m_{\chi}$) regions excluded by
relevant 
experiments \protect\cite{sigraciela}. All 
regions are defined at the 90\% confidence level. Inset: PPC spectrum 
used for the extraction of present limits. Lines display the  
signals expected from some reference WIMP candidates (dotted:
$m_{\chi}$= 8 
GeV/c$^{2}$, $\sigma_{SI}=10^{-4}$pb. Dashed: $m_{\chi}$= 6 
GeV/c$^{2}$, $\sigma_{SI}=0.002$ pb. Dash-dotted: $m_{\chi}$= 4 
GeV/c$^{2}$, $\sigma_{SI}=10^{-2}$pb).
}
\end{figure}

The signal from the PPC 
preamplifier is sent through two shaping amplifiers operating
at different integration constants. 
An anomalous ratio between the amplitudes of 
these shaped pulses is an efficient tag for microphonic events \cite{morales}. These
software cuts, 
applied on the digitized and stored amplifier traces, are trained
on datasets consisting of asymptomatic low-energy signals from an 
electronic pulser. The goal is 
to obtain the maximum signal acceptance 
for the best possible microphonic rejection. A correction
is also applied 
to the data, to compensate for the
modest signal acceptance 
losses (few percent) 
imposed by this method. The energy resolution and calibration were
obtained 
using 
the  cosmogenic activation in 
$^{71}$Ge ($T_{1/2}$=11.4 d), leading to intense 
peaks at 1.29 keV and 10.36 keV following 
installation, and a $^{133}$Ba source 
providing five auxiliary lines below 400 keV. An 
excellent linearity was observed. The energy resolution $\sigma$
below 10 keV 
is approximated by 
$\sigma^{2}=\sigma_{n}^{2}+ (2.35)^{2} E \eta F$, where $\sigma_{n}$=69.7 eV is 
the intrinsic electronic noise, $E$ is the energy in eV, $\eta$= 2.96
eV is the average energy required to create an 
electron-hole pair in Ge at $\sim$80 K, and $F\sim$0.06 is the 
measured Fano factor.

The spectrum of energy depositions 
so obtained can then be compared with expected signals from 
a standard isothermal
galactic WIMP halo. The spectrum of 
WIMP-induced recoil energies is 
generated following \cite{smith}, using a local WIMP density of 
$0.3$ GeV/cm$^{3}$, a halo velocity dispersion of 230 km/s, an
Earth-halo 
velocity of 244 km/s and a galactic escape velocity of 650 km/s. The 
quenching factor (i.e., the fraction of recoil energy measurable as 
ionization) for sub-keV germanium recoils has been 
measured with this PPC, using
a dedicated 24 keV neutron beam \cite{kansas}. It was  
found to be in 
excellent agreement with expectations 
\cite{jcap,inpreparation}. Its effect is 
included here in generating spectral shapes
of WIMP-induced ionization or ``electron 
equivalent'' energy (units of ``keVee''),
like those shown in the inset of {\frenchspacing Fig. 2}. The
exceptional energy 
resolution of this detector
has a negligible effect on 
these spectra. A standard method \cite{science,etc} can 
then be used to obtain limits on the maximum 
WIMP signal compatible with the data: 
employing a non-linear regression algorithm, data are fitted 
by a model consisting of {\it i}) 
a simple exponential to represent the spectral 
shape of low-energy backgrounds, {\it ii}) a gaussian
peak at 1.29 keV ($^{68,71}$Ge) with free 
amplitude and a width (resolution) as described 
above, and {\it iii}) for WIMPs of each mass, their 
spectral shape with a free 
normalization proportional to the spin-independent 
WIMP-nucleus coupling. Couplings excluded at the 90\% confidence 
level are plotted in {\frenchspacing Fig. 2}. The last 
remaining region of phase space available for a standard isothermal
WIMP halo 
to be the source of the DAMA modulation is now ruled out. Other more 
elaborate  
halo models might be invoked, but they result in a modest distortion
of experimental exclusion lines 
and DAMA favored phase space, both following a similar 
displacement within {\frenchspacing Fig. 2} \cite{sigraciela}. Channeling through crystal 
lattices has been proposed \cite{chakat,chadro,chadama,chafran} as a mechanism able to 
recover the compatibility of DAMA and other experiments, 
even if experimental evidence in the relevant 
recoil energy regime for NaI[Tl] seems absent \cite{chaina}. HPGe  
should also be subject to this presumptive 
effect \cite{chadama}, leading again to an expected analogous 
drift of DAMA region and PPC exclusions in {\frenchspacing Fig. 2}. 
Reference \cite{chakat} does not include a 
calculation of channeling for HPGe.
In addition, it seems unlikely that
compatibility could be recovered in these more {\it ad hoc} scenarios. 

\begin{figure}
\includegraphics[width=7.5cm]{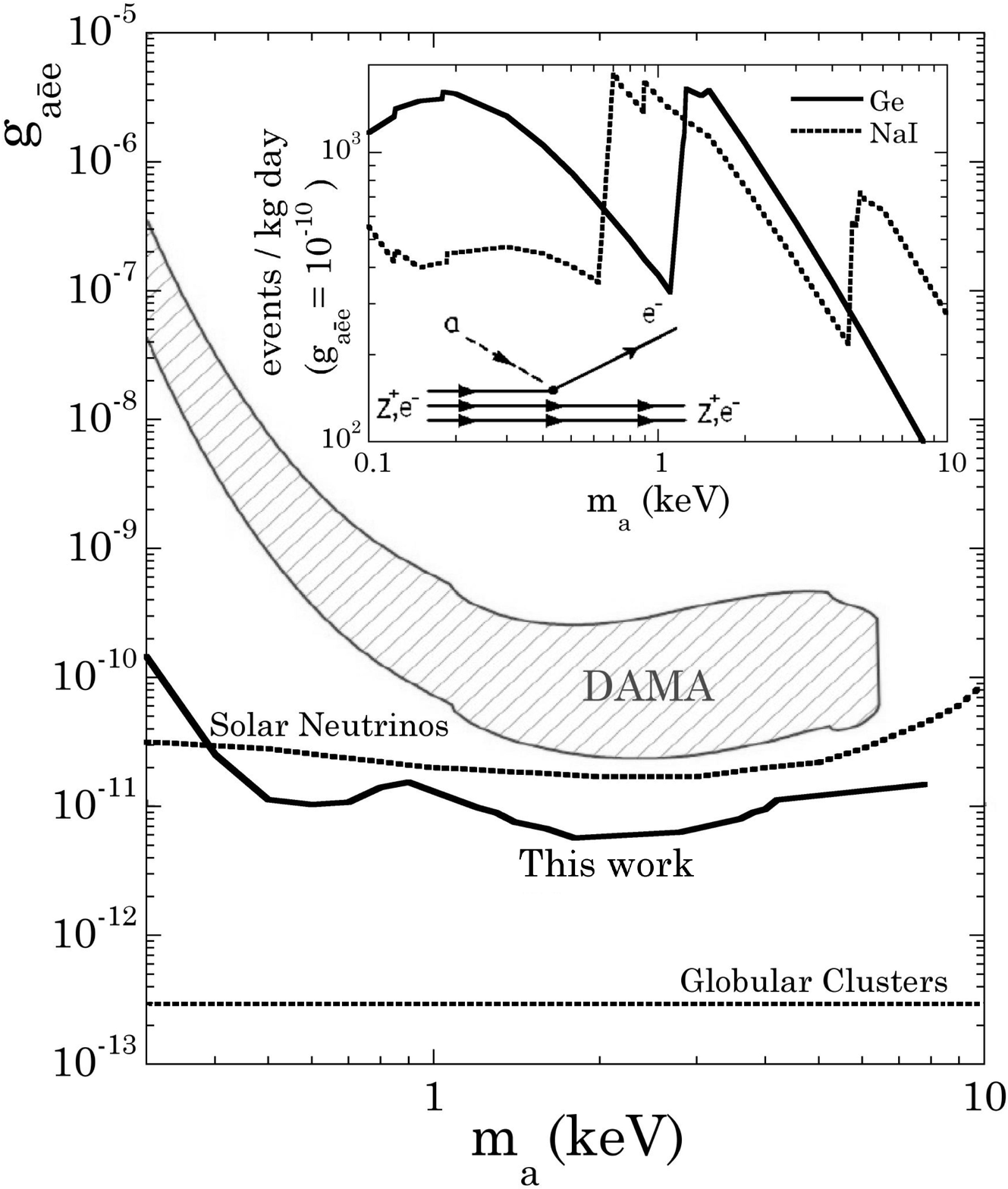}
\caption{\label{fig:epsart} Hatched region: viable parameter space in
an interpretation of the DAMA modulation involving an axio-electric 
coupling $g_{a\bar e e}$ from pseudoscalars composing a dark
isothermal 
halo, according to
\protect\cite{damapseudo}. The validity of this interpretation is now 
challenged \protect\cite{maxim} (see text). The solid line indicates 
present limits, dashed lines recent astrophysical bounds 
\protect\cite{georg}.
Inset: expected pseudoscalar interaction rates 
in Ge and NaI, for a fixed value of $g_{a\bar e e}$, as a function of 
pseudoscalar mass $m_{a}$. }
\end{figure}

While the WIMP hypothesis may at this point seem an unlikely
explanation to
the DAMA modulation, the DAMA collaboration has 
reminded us that dark matter candidates are numerous 
\cite{damaclaim,damalight,damapseudo}. Of these, axion-like dark 
pseudoscalars are arguably comparable to WIMPs in their naturalness,
being the subject of many dedicated searches. It has been 
claimed \cite{damapseudo} that such a pseudoscalar, coupling to 
electrons via the axio-electric effect, might be responsible for the 
observed modulation. Following the prescriptions in \cite{smith} 
and the proportionality between 
axio-electric and photo-electric couplings described in \cite{frank},
it would be 
possible to arrive at a compact expression for the axio-electric 
interaction rate 
from pseudoscalars forming a standard dark halo with the properties 
listed above, acting on a target of mass number $A$. However, the cross 
section in \cite{frank} tacitly assumes relativistic 
particle speeds, not the case here. The correct 
expression for the relevant interaction rate is derived in \cite{maxim}: 
$R$ [kg$^{-1}$d$^{-1}$] = $1.2\times10^{19} A^{-1} ~ g^{2}_{a\bar e 
e} ~ m_{a} ~ \sigma_{pe}$, where $g_{a\bar e 
e}$ is the dimensionless strength of the coupling, 
$m_{a}$ is the pseudoscalar rest mass in keV and $\sigma_{pe}$ is the 
photo-electric cross section in barns/atom. These rates are shown 
for both NaI and Ge in the inset of {\frenchspacing Fig. 3.} Due to
the non-relativistic nature of galaxy-bound dark matter, the spectral 
observable from such interactions would be a peak at an energy 
corresponding to $m_{a}$. DAMA actually observes the bulk of the 
modulation being centered around such a peak 
at $\sim$3 keV \cite{damaclaim}, albeit 
hindered by another one from a known source of radioactive
contamination 
($^{40}$K). Using a non-linear fitting algorithm and exponential
background 
model as above, it is possible to place 
90\% C.L. limits on the maximum amplitude of a gaussian peak of width
defined by
the energy resolution of the detector, buried 
anywhere 
in the 0.3-8 keV PPC spectral region. The rate under this peak
is then correlated to an 
excluded value of $g_{a\bar e 
e}$ via the expression above. These constraints are displayed in 
{\frenchspacing Fig. 3} together with the values of $g_{a\bar e 
e}$ and
$m_{a}$ claimed to be compatible with the DAMA effect in \cite{damapseudo}. While the 
cosmological relevance of pseudoscalar, scalar and vector dark matter in the keV 
mass region is emphasized in \cite{maxim}, it is also shown there 
that the DAMA modulation is too large to be caused by these 
possibilities. 
We therefore caution the reader about the relevance of 
the DAMA region in {\frenchspacing Fig. 3} and of the 
conclusions in \cite{damapseudo}. More specifically, adopting the rates and 
reasoning in \cite{maxim}, a pseudoscalar origin for the modulation 
cannot be justified, but DAMA should still have a competitive sensitivity 
to such candidates.

\begin{figure}
\includegraphics[width=7.7cm]{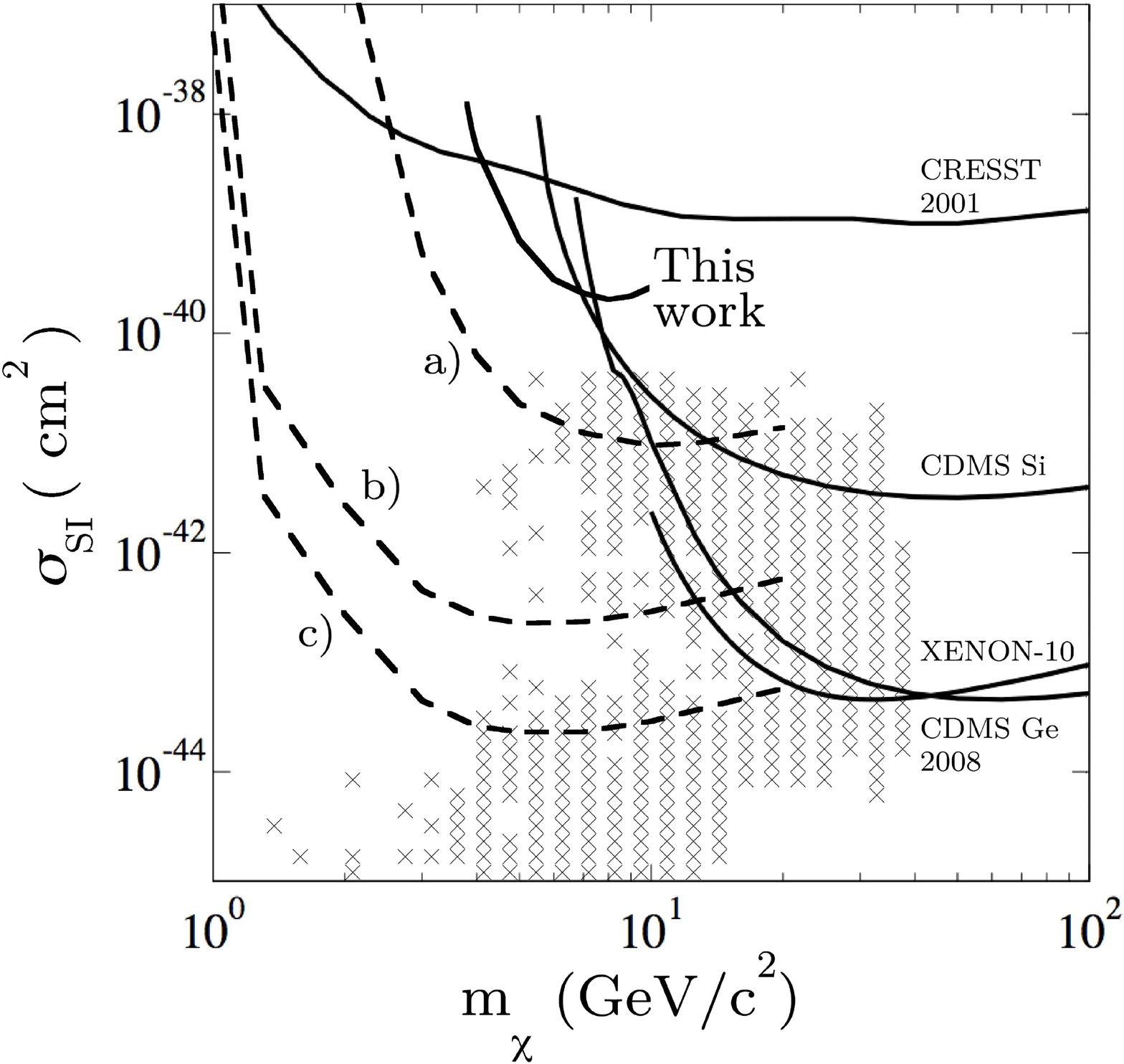}
\caption{\label{fig:epsart}  Solid lines: spin-independent 
sensitivity from leading experiments in the  
WIMP low-mass region. 
A theoretically-favored NMSSM phase space is denoted by crosses. 
Dashed lines: predicted sensitivity for PPC HPGe in a 
number of scenarios \protect\cite{jcap}: a) expected
reduction in background 
from cryostat upgrade, b) background reduction to best achieved 
 in HPGe \protect\cite{igexdm} plus an improvement to 100 eV 
 threshold, c) very conservative limiting sensitivity 
imposed by $\sim$15 d of cosmogenic $^{3}$H production at 
sea level for the {\sc{Majo\-ra\-na}} demonstrator array (a best estimate 
represents 
a $\times10$ further improvement).}
\end{figure}

An effort is in progress to further reduce the electronic 
noise in PPCs \cite{jcap}: improvements to FET 
(Field Effect Transistor, the first 
amplification stage) configuration and packaging, preamplifier
design, etc., 
are under active investigation. Detectors like these, 
with a capacitance of
$\sim$1 pF, should  be capable in principle of ionization energy
thresholds below 
100 eV. The {\sc{Majo\-ra\-na}} collaboration plans to experiment
with a 
$\sim$40 kg target mass of PPCs as part of a 
60 kg demonstrator array, to profit from their enhanced gamma background 
rejection \cite{jcap}. It is natural to wonder about the
possible 
reach of {\sc{Majo\-ra\-na}} PPCs as dark matter detectors, and
specifically 
about particle phenomenologies where all other existing 
detector designs would be unable to contribute to the 
exploration, due to
their higher thresholds.

Several scenarios 
have been proposed where naturally light ($<$10 GeV/c$^{2}$) WIMPs 
appear \cite{maxim,bottinog,stephano,minscalar}. Q-balls can similarly lead to modest ionization signals
\cite{alex}. The lightest neutralino, an electrically neutral
particle present in supersymmetric extensions of the Standard Model
(SM), is a well motivated candidate for WIMP dark matter
\cite{Jungman:1995df}. Its properties have been studied mostly within
the Minimal 
Supersymmetric Standard Model (MSSM) \cite{munoz}, 
where very light neutralinos
with large detection cross sections
were found to be possible \cite{vlmssm}. 
The  Next-to-Minimal Supersymmetric Standard Model
(NMSSM) is a well-justified extension of the MSSM which 
elegantly generates a Higgsino mass parameter of electroweak 
scale 
through the introduction of a new chiral singlet superfield. 
This has
interesting implications for neutralino 
dark matter \cite{nmssmdm}, and 
new regions of the parameter space exist which lead to light
neutralinos with the correct dark matter relic density
\cite{vlnmssm}. In order to illustrate these properties various scans
of the NMSSM parameter space have been performed with the code
NMHDECAY
\cite{nmhdecay}.   The 
choice of input parameters is 
beyond the scope of this letter and will be given in 
\cite{david}. The favored space is shown by crosses 
in {\frenchspacing Fig. 4}. A conservative projected sensitivity 
for {\sc{Majo\-ra\-na}} PPCs is also displayed. 
A clear complementarity to 
other detection schemes is observed.

In conclusion, by virtue of their sensitivity to 
very small energy depositions, large mass and excellent energy
resolution, 
PPC detectors are ideally suited 
for confirming or definitively disproving DAMA's claim of dark matter 
discovery. Clearly, technologies able to explore the many possible 
phenomenological faces of the dark matter problem should be 
encouraged and 
developed. The unresolved mystery of the DAMA annual modulation is a 
reminder of how often surprises arise in particle physics, where and 
when they are least expected.

\section{ACKNOWLEDGMENTS}
We are indebted to M. Pospelov for pointing out the correct speed dependence of 
the pseudoscalar cross-section, and to G. Raffelt for ensuing 
discussions. Our gratitude also goes to Tom Economou and to the Metropolitan Water
Reclamation 
District of Greater Chicago for allowing access to the TARP 
underground facilities. 
This work was supported by NSF grant PHY-0653605, 
NSF CAREER award PHY-0239812, NNSA 
grant DE-FG52-0-5NA25686, 
the Kavli Institute for Cosmological Physics through NSF grant  
PHY-0114422, Argonne and Pacific Northwest National Laboratory LDRD 
funds, Juan de la Cierva program of the Spanish MEC and by the 
Comunidad de Madrid under project HEPHACOS.
We would like to thank G. Gelmini, P. Gondolo, A. 
Kusenko and our colleagues within the 
{\sc{Majo\-ra\-na}} collaboration for many useful exchanges.

\end{document}